\documentclass[sigconf,nonacm,screen]{acmart}
\acmConference[ESEC/FSE 2021]{The 29th ACM Joint European Software Engineering Conference and Symposium on the Foundations of Software Engineering}{23 - 27 August, 2021}{Athens, Greece}

\usepackage{multirow}
\usepackage{tabularx}
\usepackage{colortbl}
\usepackage{enumitem,kantlipsum}

\AtBeginDocument{%
  \providecommand\BibTeX{{%
    \normalfont B\kern-0.5em{\scshape i\kern-0.25em b}\kern-0.8em\TeX}}}

\newcommand{\tabincell}[2]{\begin{tabular}{@{}#1@{}}#2\end{tabular}}

\begin{document}

\title{CodeXGLUE: A Machine Learning Benchmark Dataset for Code Understanding and Generation}

\author{Shuai Lu}
\authornote{indicates equal contribution and internship at Microsoft. Authors are listed in alpha-beta order. Corresponding authors are Duyu Tang and Shujie Liu.}
\affiliation{%
  \institution{Peking University}
  }

\author{Daya Guo}
\authornotemark[1]
\affiliation{%
  \institution{Sun Yat-sen University}
  }

\author{Shuo Ren}
\authornotemark[1]
\affiliation{%
  \institution{Beihang University}
  }

\author{Junjie Huang}
\authornotemark[1]
\affiliation{%
  \institution{Beihang University}
  }

\author{Alexey Svyatkovskiy}
\affiliation{%
  \institution{Microsoft}
  }

\author{Ambrosio Blanco}
\affiliation{%
  \institution{Microsoft Research Asia}
  }

\author{Colin Clement}
\affiliation{%
  \institution{Microsoft}
  }

\author{Dawn Drain}
\affiliation{%
  \institution{Microsoft}
  }

\author{Daxin Jiang}
\affiliation{%
  \institution{Microsoft}
  }

\author{Duyu Tang}
\affiliation{%
  \institution{Microsoft Research Asia}
  }

\author{Ge Li}
\affiliation{%
  \institution{Peking University}
  }

\author{Lidong Zhou}
\affiliation{%
  \institution{Microsoft Research Asia}
  }

\author{Linjun Shou}
\affiliation{%
  \institution{Microsoft}
  }

\author{Long Zhou}
\affiliation{%
  \institution{Microsoft Research Asia}
  }

\author{Michele Tufano}
\affiliation{%
  \institution{Microsoft}
  }

\author{Ming Gong}
\affiliation{%
  \institution{Microsoft}
  }

\author{Ming Zhou}
\affiliation{%
  \institution{Microsoft Research Asia}
  }

\author{Nan Duan}
\affiliation{%
  \institution{Microsoft Research Asia}
  }

\author{Neel Sundaresan}
\affiliation{%
  \institution{Microsoft}
  }

\author{Shao Kun Deng}
\affiliation{%
  \institution{Microsoft}
  }

\author{Shengyu Fu}
\affiliation{%
  \institution{Microsoft}
  }

\author{Shujie Liu}
\affiliation{%
  \institution{Microsoft Research Asia}
  }

\renewcommand{\shortauthors}{Lu, Guo, Ren and Huang, et al.}

\begin{abstract}
Benchmark datasets have a significant impact on accelerating research in programming language tasks. 
In this paper, we introduce CodeXGLUE, a benchmark dataset to foster machine learning research for program understanding and generation. CodeXGLUE includes a collection of 10 tasks across 14 datasets and a platform for model evaluation and comparison. CodeXGLUE also features three baseline systems, including the BERT-style, GPT-style, and Encoder-Decoder models, to make it easy for researchers to use the platform.
The availability of such data and baselines can help the development and validation of new methods that can be applied to various program understanding and generation problems
\footnote{CodeXGLUE is publicly available at \url{https://github.com/microsoft/CodeXGLUE}. Participants can submit their results by emailing to \url{codexglue@microsoft.com}.}.
\end{abstract}




\maketitle

\section{Introduction}

Evans Data Corporation\footnote{\url{https://evansdata.com/press/viewRelease.php?pressID=278}} estimated that there were 23.9 million professional developers in 2019 and that the number was expected to reach 28.7 million in 2024. With the population of developers growing at such a rate, 
 code intelligence that 
leverages artificial intelligence (AI) to help software developers improve the productivity of the development process is becoming increasingly important.
It is commonly accepted that 
%
benchmarks have a significant impact  on the growth of applied AI research.
In this paper, we focus on establishing a benchmark dataset for code intelligence.

Automated program understanding and generation could increase the productivity of software developers.
In fact, developers who want to find code written by others with the same intent can leverage code search systems \cite{husain2019codesearchnet,gu2018dcs,wang2020trans,premtoon2020} to automatically retrieve semantically relevant codes through natural language queries. Similarly, developers who are confused about what to write next can use code completion systems \cite{raychev2016probabilistic,allamanis2013mining,raychev2014completion,svyatkovskiy2019pythia,svyatkovskiy2020intellicode,bruch2009,hindle2012naturalness,bielik2016phog} to automatically complete the following tokens based on the edits made to the code. Finally, when developers want to implement the Java code in Python, code-to-code translation systems \cite{nguyen2015divide,karaivanov2014,chen2018tree,lachaux2020unsupervised} can help translate their code from one programming language (Python) to another (Java).

In recent years,  researchers have increasingly applied statistical models, including neural nets, to code intelligence tasks. Very recently, the application of pretrained models that learn from big programming language data has been inspired by the great success of pretrained models like BERT \cite{devlin2018bert} and GPT \cite{solaiman2019release} in natural language processing (NLP). These models, including CodeBERT \cite{feng2020codebert} and IntelliCode Compose \cite{svyatkovskiy2020intellicode}, have led to further improvements in code understanding and generation problems, but they 
lack a benchmark suite that covers a wide range of tasks. 
The use of ImageNet  \cite{deng2009imagenet} for computer vision and the use of GLUE \cite{wang2018glue} for NLP have shown that a diversified benchmark dataset has a significant impact on the growth of applied AI research.

\newcolumntype{x}{>{\columncolor[rgb]{0.52,0.80,0.97}}c}
\begin{table*}[h]
    \setlength\aboverulesep{0pt}
    \setlength\belowrulesep{0pt}
    \begin{small}
	\caption{A brief summary of CodeXGLUE, which includes tasks, datasets, languages, sizes in various states, and baseline systems. Highlighted datasets are newly introduced.}
	\label{table-tasks}
	\begin{center}
		\begin{tabular}{cxxxxc}
			\toprule 
			\rowcolor{white}
			Category & Task & Dataset Name & Language & Train/Dev/Test Size & Baselines \\
			\midrule 
			\rowcolor{white}
			\multirow{9}{*}[-7ex]{Code-Code} & & BigCloneBench \cite{svajlenko2014towards} & Java & 900K/416K/416K & \\
			\cmidrule{3-5}
			\rowcolor{white}
			& \multirow{2}{*}[3ex]{Clone Detection} & POJ-104 \cite{mou2016convolutional} & C/C++ & 32K/8K/12K & \\
			\cmidrule{2-5}
			\rowcolor{white}
			& Defect Detection & Devign \cite{zhou2019devign} & C & 21K/2.7K/2.7K & \\
			\cmidrule{2-5}
			& &  CT-all &  \tabincell{c}{Python,Java,PHP,\\JavaScript,Ruby,Go} &  -/-/176K & \\
			\cmidrule{3-5}
			 & \multirow{2}{*}[4.5ex]{Cloze Test} &  CT-max/min \cite{feng2020codebert} & \tabincell{c}{Python,Java,PHP,\\JavaScript,Ruby,Go} & -/-/2.6K & \multirow{5}{*}[15ex]{CodeBERT}\\
			\cmidrule{2-6}
			& & PY150 \cite{raychev2016probabilistic} & Python & 100K/5K/50K & \multirow{2}{*}[-1ex]{CodeGPT} \\
			\cmidrule{3-5}
			& \multirow{2}{*}[3ex]{Code Completion} & Github Java Corpus\cite{allamanis2013mining} & Java & 13K/7K/8K & \\
			\cmidrule{2-6}
			\rowcolor{white}
			& Code Repair & Bugs2Fix \cite{tufano2019empirical} & Java & 98K/12K/12K & \multirow{2}{*}[-0.5ex]{\tabincell{c}{Encoder-\\Decoder}} \\
			\cmidrule{2-5}
			& Code Translation & CodeTrans & Java-C\# & 10K/0.5K/1K & \\
			\midrule
			\multirow{3}{*}[-5ex]{Text-Code} & & \tabincell{c}{CodeSearchNet \cite{husain2019codesearchnet},\\AdvTest} & Python & 251K/9.6K/19K & \multirow{2}{*}[-2ex]{CodeBERT} \\
			\cmidrule{3-5}
			& \multirow{2}{*}[4.5ex]{NL Code Search} & \tabincell{c}{CodeSearchNet \cite{husain2019codesearchnet},\\WebQueryTest} & Python & 251K/9.6K/1K & \\
			\cmidrule{2-6}
			\rowcolor{white}
			& \tabincell{c}{Text-to-Code\\Generation} & CONCODE \cite{iyer2018mapping} & Java & 100K/2K/2K & CodeGPT \\
			\midrule
			\rowcolor{white}
			Code-Text & Code Summarization & CodeSearchNet \cite{husain2019codesearchnet} & \tabincell{c}{Python,Java,PHP,\\JavaScript,Ruby,Go} & 908K/45K/53K & \multirow{2}{*}[-2ex]{\tabincell{c}{Encoder-\\Decoder}} \\
			\cmidrule{1-5}
			Text-Text & \tabincell{c}{Documentation\\Translation} & Microsoft Docs & \tabincell{c}{English-Latvian/Danish\\/Norwegian/Chinese} & 156K/4K/4K & \\
			\bottomrule
		\end{tabular}
	\end{center}
	\end{small}
\end{table*}


To address this problem, we introduce CodeXGLUE, a machine learning benchmark dataset for program understanding and generation research that  includes 14 datasets\footnote{We plan to evolve the benchmark over time by extending to more tasks. }, a collection of 10 diversified programming language understanding and generation tasks, and a platform for model evaluation and comparison. CodeXGLUE supports the following tasks:

\begin{itemize}
    \item \textbf{code-code} (clone detection \cite{svajlenko2014towards,white2016deep,mou2016convolutional,wang2020detecting,buch2019clone,zhang2019novel,ye2020misim}, defect detection \cite{zhou2019devign,ray2016naturalness,michael2018deepbugs,li2019improving,wang2016bugram,wang2016auto}, cloze test \cite{feng2020codebert}, code completion \cite{raychev2016probabilistic,allamanis2013mining,raychev2014completion,svyatkovskiy2019pythia,svyatkovskiy2020intellicode,bruch2009,hindle2012naturalness,bielik2016phog}, code repair \cite{tufano2019empirical,allamanis2017learning,gupta2017deepfix,hellendoorn2019global,vasic2019neural,vekris2016refine}, and code-to-code translation \cite{nguyen2015divide,karaivanov2014,chen2018tree,lachaux2020unsupervised})
\item \textbf{text-code} (natural language code search \cite{husain2019codesearchnet,gu2018dcs,wang2020trans}, text-to-code generation \cite{iyer2018mapping,clement2020pymt5,yin2017syntactic,wei2019code,xu2020incorporating,yin2018mining,iyer2019learning,guo2019coupling})
\item \textbf{code-text} (code summarization \cite{iyer2016summarizing,clement2020pymt5,fernandes2018structured,wan2018improving,wang2020trans,wei2019code,allamanis2016convolutional,hu2018sum,Wang2020CoCoGUMCC})
\item \textbf{text-text} (documentation translation \cite{johnson2017google})
\end{itemize}


CodeXGLUE includes eight previously proposed datasets — BigCloneBench \cite{svajlenko2014towards}, POJ-104 \cite{mou2016convolutional}, Devign \cite{zhou2019devign}, PY150 \cite{raychev2016probabilistic}, Github Java Corpus \cite{allamanis2013mining}, Bugs2Fix \cite{tufano2019empirical}, CONCODE \cite{iyer2018mapping}, and CodeSearchNet \cite{husain2019codesearchnet}— but also newly introduced datasets that are highlighted in Table \ref{table-tasks}. 
The datasets are chosen or created based on the consideration that the task has clear definition, and the volume of the dataset could support the development and evaluation of data-driven machine learning methods.
The datasets created by us include (1) two cloze test test sets that cover 6 programming languages, (2) two line-level code completion test sets in Java and Python, respectively, (3) a code-to-code translation dataset between Java and C\#, (4) two natural language code search test sets with web queries and normalized function and variable names, respectively, and (5) a documentation translation dataset that covers five natural languages.

To make it easy for participants, we provide three baseline models to help perform the tasks, including a BERT-style pretrained model (in this case, CodeBERT) to supports code understanding problems, a GPT-style pretrained model, which we call CodeGPT, to help solve completion and generation problems, and an Encoder-Decoder framework that tackles sequence-to-sequence generation problems. 

\section{Tasks Overview}
In this section, we provide a definition for each task. 
\begin{itemize}
[wide, labelwidth=!, labelindent=5pt,label={}]
    \item \textbf{Clone detection \cite{svajlenko2014towards,mou2016convolutional}.} The task is to measure the semantic similarity between codes. This includes two subtasks: binary classification between a pair of codes and code retrieval, where the goal is to find semantically similar codes.
    \item \textbf{Defect detection \cite{zhou2019devign}}. The objective is to  identify whether a body of source code contains defects that may be used to attack software systems, such as resource leaks, use-after-free vulnerabilities, and DoS attack. 
    \item \textbf{Cloze test \cite{feng2020codebert}}. This aims to predict the masked token of a code and includes two subtasks. The first one is to measure the accuracy of 
    predicting the
    masked token from the whole vocabulary. The other is to test the semantic reasoning ability by distinguishing between ``max'' and ``min''.
    \item \textbf{Code completion \cite{raychev2016probabilistic,allamanis2013mining}}. It aims to predict following tokens based on a code context. Its subtasks are token-level completion and line-level completion.  The former checks whether the next one token has been predicted correctly, while the latter tests the goodness of the generated line.
    \item \textbf{Code translation \cite{nguyen2015divide}}. It involves translating a code from one programming language to a different one. 
    \item \textbf{Code search \cite{husain2019codesearchnet}.} It measures the semantic relatedness between texts and codes and is composed of two subtasks. 
    The first one is to find the most relevant code in a collection of codes according to a natural language query.
    The second subtask entails the analysis of a query-code pair to predict whether the code answers the query or not. 
    \item \textbf{Code repair \cite{tufano2019empirical}}. Its goal is to  refine the code by fixing the bugs automatically. 
    \item \textbf{Text-to-code generation \cite{iyer2018mapping}}. This aims to  generate a code via a natural language description. 
    \item \textbf{Code summarization \cite{iyer2016summarizing}}. The objective is to generate the natural language comment for a code. 
    \item \textbf{Documentation translation \cite{johnson2017google}}. It aims to  translate code documentation from one natural language to different one. 
    \end{itemize}

\section{Datasets}
In this section, we describe the datasets included in CodeXGLUE. Datasets are chosen or created based on the criterion that the volume of the dataset could support the development and evaluation of data-driven machine learning methods.

\subsection{Clone detection}
Clone detection includes two subtasks. The first subtask is to predict whether two given codes have the same semantics. We use the BigCloneBench \cite{svajlenko2014towards} dataset for the subtask.
The second subtask aims to retrieve semantically similar codes given a code as the query and we use the dataset POJ-104 \cite{mou2016convolutional} to perform it.


\textbf{BigCloneBench} is a widely used large code clone benchmark that contains over 6,000,000 true clone pairs and 260,000 false clone pairs from 10 different functionalities. The dataset provided by \citet{wang2020detecting} is filtered by discarding code fragments without any tagged true or false clone pairs, leaving it with 9,134 Java code fragments. Finally, the dataset includes 901,028/415,416/415,416 examples for training, validation and testing, respectively.      

\textbf{POJ-104} dataset \cite{mou2016convolutional} comes from a pedagogical programming open judge (OJ) system that automatically judges the validity of submitted source code for specific problems by running the code. We use the POJ-104 dataset, which consists of 104 problems and includes 500 student-written C/C++ programs for each problem. Different from that of the BigCloneBench dataset, the task of POJ-104 aims to retrieve other programs that solve the same problem given a program. 
We group the datasets in three subsets based on the number of problems they are required to solve (64/16/24) for training, validation, and testing.


\subsection{Defect detection}
\label{defect_detection_dataset}
For the task of defect detection, \citet{zhou2019devign} provide the \textbf{Devign} dataset that includes 27,318 manually-labeled functions collected from two large C programming language open-source projects popular among developers and diversified in functionality, i.e., QEMU and FFmpeg. The dataset was created by collecting security-related commits and extracting vulnerable or non-vulnerable functions from the labeled commits. 
Since \citet{zhou2019devign} did not provide official training/validation/testing sets for the two projects, we randomly shuffle the dataset and split 80\%/10\%/10\% of the dataset for training/validation/testing. The task is formulated as a binary classification to predict whether a function is vulnerable.

\subsection{Cloze test}
\begin{figure}[h]
    \includegraphics[width=8.3cm]{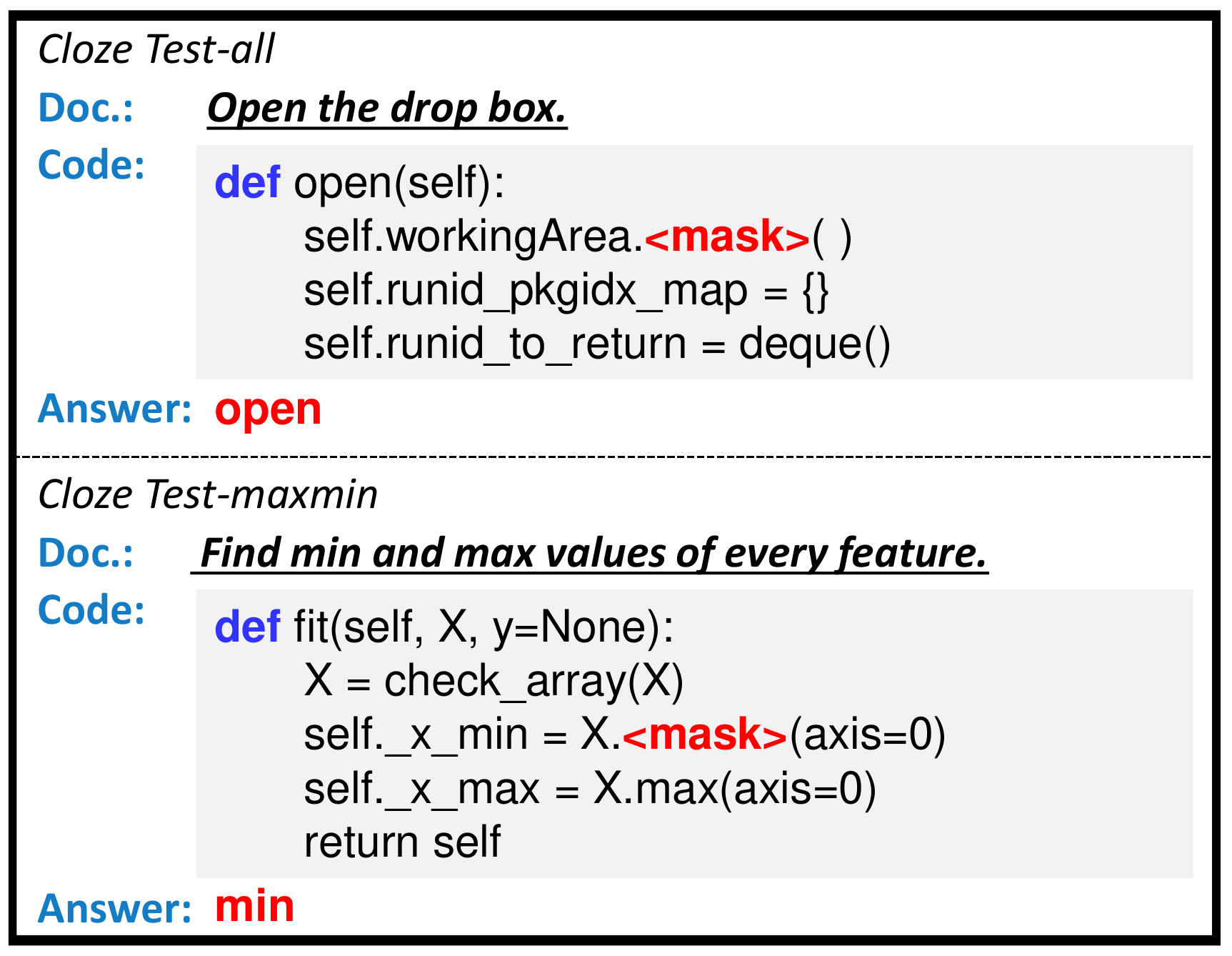}
    \caption{Two examples in the cloze test dataset.}
    \label{fig:clozetest-example}
\end{figure}

Figure \ref{fig:clozetest-example} shows two examples of the cloze test (CT) task in code domain, which aims to assess models' ability to understand a code by asking those models to predict the masked code from several candidates. 
We focus on two subtasks: CT-all with candidates from a filtered vocabulary and CT-maxmin with the candidates “max” and “min”.

We use the validation and testing sets of CodeSearchNet \cite{husain2019codesearchnet} to create CT-all and CT-maxmin datasets for six programming languages, i.e., Go, Java, JavaScript (JS), PHP, Python and Ruby. 

\paragraph{CT-all} To less introduce lengthy variable names and avoid the issue caused by the use of different tokenizers, we select target cloze words by retaining unique words after Byte Pair Encoding \cite{sennrich2016bpe}, and we remove meaningless tokens like punctuations with handcrafted rules. At last, 930 tokens are selected among six languages in total. We select codes containing the 930 tokens and manually set threshold values of token occurrence to balance the frequency of the 930 tokens in CT-all. 

\paragraph{CT-maxmin} To further evaluate models' ability to understand code semantics, we introduce CT-maxmin to test how well model can distinguish the difference between \textit{max} and \textit{min}. CT-maxmin comes from the dataset used for the PL-Probing task in CodeBERT\cite{feng2020codebert}, which includes codes containing the keywords of \textit{max} or \textit{min}.

The data statistics are listed in Table \ref{table-clozetest-data-statistic}.

\begin{table}[h]
	\caption{Data statistics about the cloze test datasets.}
	\label{table-clozetest-data-statistic}
	\begin{center}
		\begin{tabular}{lcc}
			\toprule 
			Task & CT-all & CT-maxmin \\
			\midrule 
			Go       &  25,282     &    152  \\
			Java     &  40,492     &    482  \\
			JavaScript &13,837     &    272  \\
			PHP     &   51,930     &    407  \\
			Python  &   40,137     &    1,264  \\
			Ruby    &   4,437      &    38  \\
			\midrule 
			All     &   176,115    &    2,615 \\
			\bottomrule
		\end{tabular}
	\end{center}
\end{table}


\subsection{Code completion}
We use two influential datasets for code completion, \textbf{PY150} in python and \textbf{Github Java Corpus} in Java. Both datasets can help achieve token-level code completion. 
We move further by creating two test sets for the line-level code completion task from the two corpora.
The task is to complete an unfinished line. Models should be capable of predicting code sequences of arbitrary token types and code structures. 

\begin{figure}[h]
    \includegraphics[width=8.3cm]{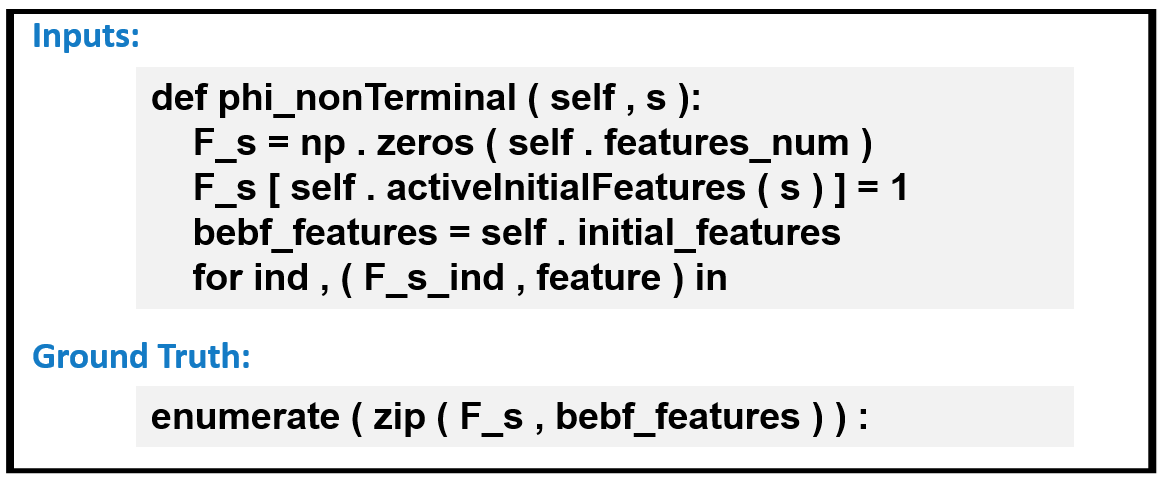}
    \caption{An example in the line-level code completion dataset.}
    \Description{A tokenized python code snippet in the format of code completion}
    \label{fig:linecompletion-sample}
\end{figure}

 \textbf{PY150} is a Python dataset \cite{raychev2016probabilistic} containing 150,000 Python source files collected from Github. We follow the data split in \citet{raychev2016probabilistic}, resulting in 100,000 files for training and 50,000 files for testing, consisting 76.3M tokens and 37.2M tokens, respectively. We preprocess the corpora by tokenizing source codes, removing comments, replacing strings with length of more than 15 characters with empty strings, and adding a special token $\langle\mathit{EOL}\rangle$ (end-of-line) to mark the ending of a line explicitly. For line-level code completion, we create 10,000 examples from different files in the test set of PY150 for testing. 
 Since we intend to test model's ability to autocomplete an arbitrary line, we select the line to be predicted at random. 
  We generate a test case by ensuring that there is sufficient context, i.e., at least 15\% of the whole file.
  Models are expected to generate the following line ended by $\langle\mathit{EOL}\rangle$ given the context.
  The average number of tokens in input and output are 489.11 and 6.56, respectively. Figure \ref{fig:linecompletion-sample} shows an example of line-level code completion. 

 \textbf{Github Java Corpus} is a Java dataset mined by \citet{allamanis2013mining}, and it collects over 14 thousand Java projects from Github. We follow the settings established by \citet{hellendoorn2017deep} as well as \citet{karampatsis2020big}, using 1\% of the subset in the corpus. We have 12,934/7,189/8,268 files for training/validation/testing, consisting of 15.8M/3.8M/5.3M tokens, respectively. We do the same preprocessing conducted on PY150, but we don't add the special token $\langle\mathit{EOL}\rangle$ since in Java the symbols ; and \} are used to mark the ending of a code statement. For line-level code completion, we create 3,000 examples for testing from different files in the test set of the corpus. 
 Similarly to the process we follow for Python, 
 the line to be predicted is selected at random from the test file.
 The average numbers of tokens are 350.62 and 10.49 in input and output, respectively.
 

\subsection{Code translation}
\label{code_translation}
The training data for code translation is the code pairs with equivalent functionality
in two programming languages. In this paper, we provide a dataset consisting of parallel codes between Java and C\#. We did not use the dataset of \citet{lachaux2020unsupervised} because they did not have the data for supervised model training.
Following \citet{nguyen2015divide} and \citet{chen2018tree}, we use the data collected from several open-source projects, i.e., Lucene\footnote{http://lucene.apache.org/}, POI\footnote{http://poi.apache.org/}, JGit\footnote{https://github.com/eclipse/jgit/}  and Antlr\footnote{https://github.com/antlr/}. We do not use Itext\footnote{http://sourceforge.net/projects/itext/} and JTS\footnote{http://sourceforge.net/projects/jts-topo-suite/} due to the license problem. 
Those projects are originally developed in Java and then ported to C\#. They are well-established systems with long developing histories and with both Java and C\# versions in use. 

The following step is to mine paired functions or methods from those projects. According to our observation, the directory structures and function or method names of the two versions are identical or similar when they are applied to the same project. Therefore, following \citet{nguyen2015divide}, we conservatively search for the functions having the same signatures in the classes with the same/similar names and included in the same/similar directory structures of both versions. We discard duplicate code pairs and the codes having multiple targets searched with the above method. After this step, we remove the pairs whose number of overlapping tokens was less than 1/3 of the sentence length. To make our data more scalable for further syntactic and semantic analysis, we also remove the functions with null function body according to their abstract syntax tree (AST). Then we build the data-flow graph \cite{guo2020graphcodebert} for each function, which represents the dependency between two variables and provides valuable semantic information for code understanding. 
Finally, a function with no data-flow extracted from the AST of a specific function is also discarded.

\begin{figure}
    \includegraphics[width=8.3cm]{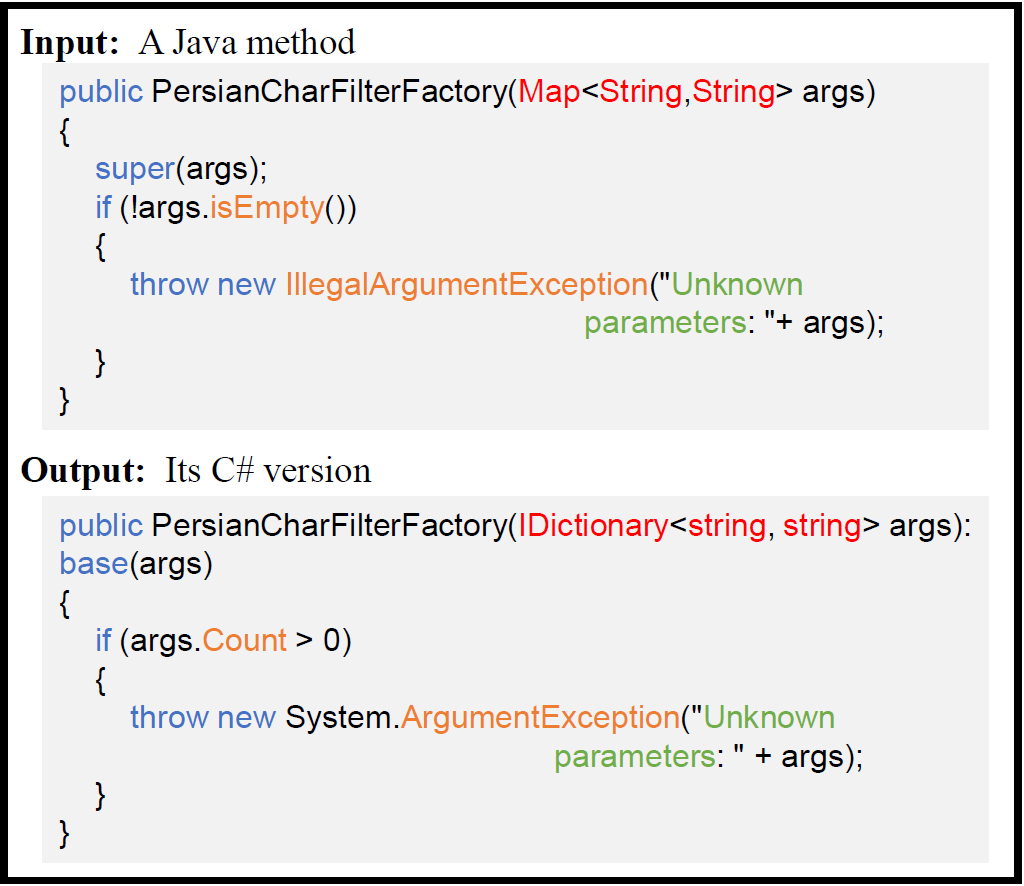}
    \caption{An example in the code translation dataset.}
    \label{fig:codetrans-example}
\end{figure}

At last, the total number of paired functions or methods is 11,800. We randomly select 500 pairs of functions for the development set and another 1,000 pairs for the test set. The average lengths of the Java and C\# functions after tokenization are 38.51 and 46.16, respectively \footnote{https://github.com/c2nes/javalang}. An example of the mined translation pairs from C\# to Java is shown in Figure \ref{fig:codetrans-example}.

\subsection{Code search}
\label{code_search}
\begin{figure}[h]
    \includegraphics[width=8.3cm]{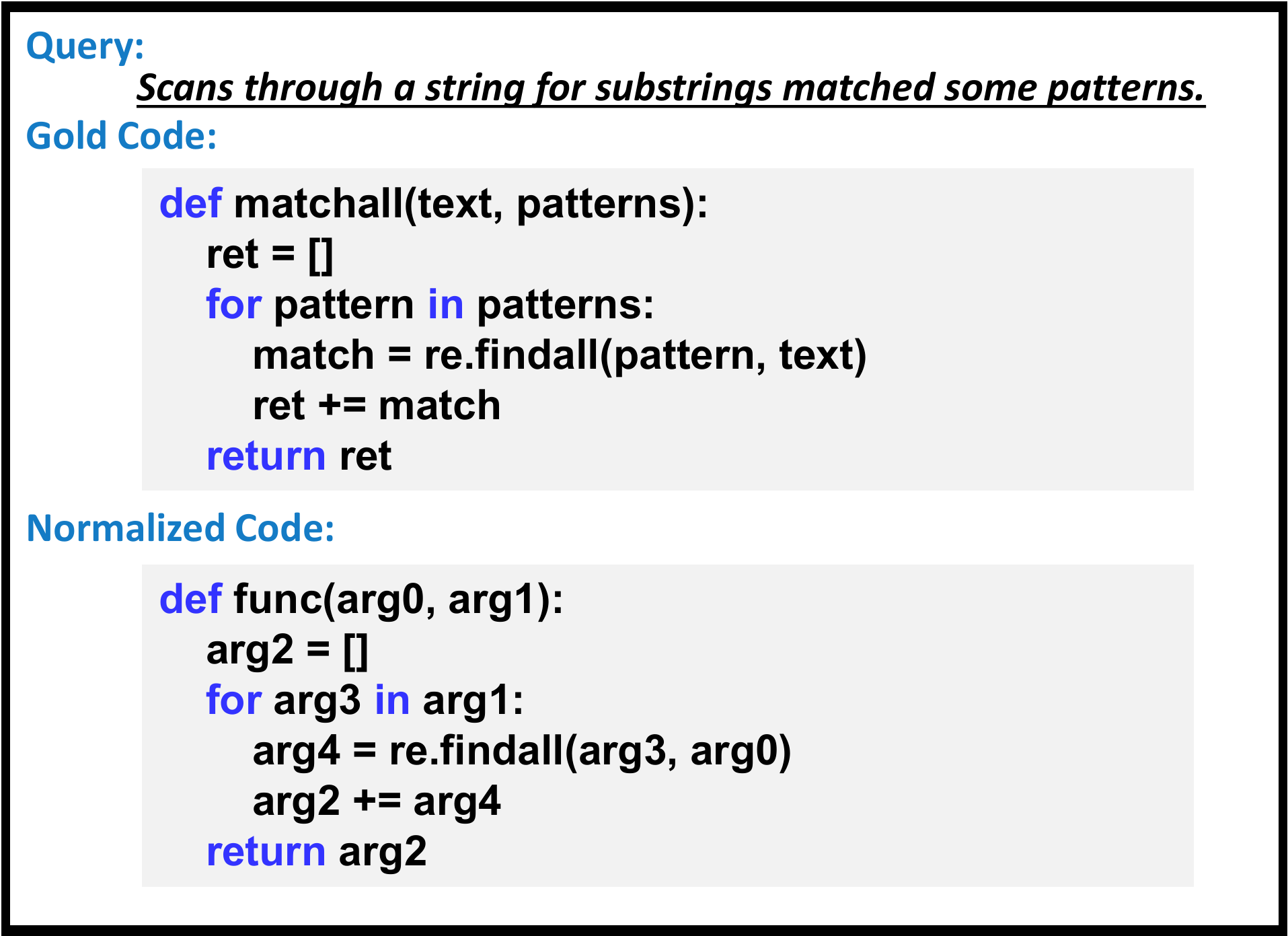}
    \caption{An example in the CodeSearchNet AdvTest dataset.}
    \label{fig:advTest-example}
\end{figure}

\begin{figure}
    \includegraphics[width=8.3cm]{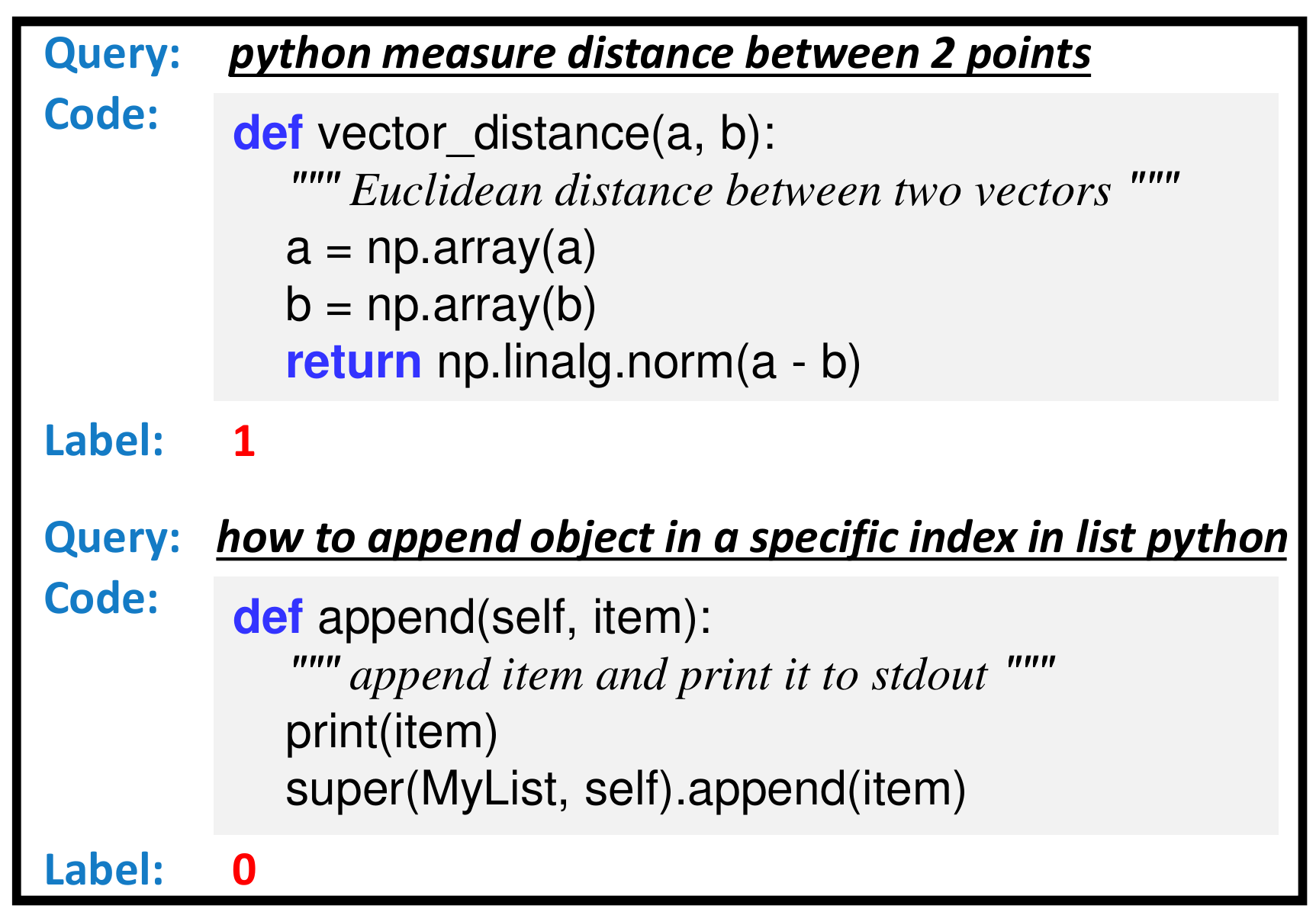}
    \caption{Two examples in the WebQueryTest dataset.}
    \label{fig:webqueryexample}
\end{figure}

Code search includes two subtasks. The first one is to find the most relevant code from a collection of candidates given a natural language query. We create a challenging testing set, called \textbf{CodeSearchNet AdvTest}, from CodeSearchNet corpus \cite{husain2019codesearchnet} for performing this task. An example of this dataset is shown in Figure \ref{fig:advTest-example}. The second subtask is to predict whether a code answers a given query.  We provide a testing set \textbf{WebQueryTest} of real user queries. Two examples of the dataset are illustrated in Figure \ref{fig:webqueryexample}.

\textbf{CodeSearchNet AdvTest} is a Python dataset from the CodeSearchNet \cite{husain2019codesearchnet} corpus. Each example includes a function paired with a document. We follow \citet{husain2019codesearchnet} to take the first paragraph of the documentation as the query for the corresponding function. To improve the quality of the dataset, we filter it by removing the following examples. 

(1) Examples whose code could not be parsed into abstract syntax tree.

(2) Examples whose document tokens number is  shorter than 3 or larger than 256.

(3) Examples whose document contains special tokens such as ``http://".

(4) Examples whose document is empty or not written in English.

At the end of the process, we obtain a dataset with 251,820 / 9,604 / 19,210 examples for training/validation/testing. 
After normalizing function or variable names with special tokens, we observe that the Mean Reciprocal Rank (MRR) scores of RoBERTa \cite{liu2019roberta} and CodeBERT \cite{feng2020codebert} for the code search task on the CodesearchNet \cite{husain2019codesearchnet} dataset drop from 0.809 to 0.419 and from 0.869 to 0.507, respectively, in Python programming language.
To better test the understanding and generalization abilities of the model, we normalize function and variable names in testing and development sets like $func$ for the function name and $arg_i$ for the i-th variable name.
Figure \ref{fig:advTest-example} shows an example in CodeSearchNet AdvTest dataset.
The task aims to search source codes from candidates for a natural language query. 
In contrast to the testing phase of previous works \cite{husain2019codesearchnet,feng2020codebert} that only involved 1,000 candidates, we use the entire testing set for each query, which makes \textbf{CodeSearchNet AdvTest} dataset more difficult. The training set for this task comes from the filtered CodeSearchNet dataset \cite{husain2019codesearchnet}.

\textbf{WebQueryTest}: Most code search datasets use code documentations or questions from online communities for software developers as queries, but these are  different from real user search queries. To fix this discrepancy, we provide WebQueryTest, a testing set of real code search for Python. The problem is formulated as a binary classification task  and as a complementary setting to the retrieval scenario. Given a pair of query and code function, a model needs to classify whether the code function can answer the query or not. 

The data creation process can be divided into two stages: data collection and annotation.
We first collect real user queries from the web query logs of a commercial search engine and we keep the queries with ``python''. 
Inspired by \citet{yan2020searchintent}, we design some heuristics based on keyword  exact matching to filter out queries without the code search intent.
Then we select candidate codes for each query from the Python validation and testing sets in CodeSearchNet. To shrink the candidates to be annotated for each query, we select the top two functions with the highest query-code similarity computed by a CodeBERT-based code retrieval model, which is trained on 148K automated-minded Python Stack Overflow Question-Code (StaQC) \cite{yao2018staqc} with the default parameters provided by \citet{feng2020codebert}. 

We use a two-stage annotation schema to label each instance. The first step is to judge whether the query has a code-search intent. Instances labeled as "-1" are those without code search intent. The second step is to assess whether the code (with its documentation) can answer the query. 
Instances labeled as "1" are those where the code can answer the query. Otherwise, they are labeled as  ``0''. Two examples are illustrated in Figure \ref{fig:webqueryexample}. We invite 13 developers proficient in Python to label 1,300 instances, with each annotator dealing with 100 of them. Discussions are allowed during annotation. Finally, the numbers of instances with labels -1, 0 and 1 are 254, 642 and 422, respectively. Since we are more interested in query-code matching, we include only the categories 0 and 1 in our final test set. 
The training and validation sets we use for this task are from the original CodeSearchNet dataset \cite{husain2019codesearchnet}.



\subsection{Code repair}
\label{code_repair}
Code repair aims to fix bugs in the code automatically. We use the dataset released by \citet{tufano2019empirical}. The source is buggy Java functions, whereas the target is the corresponding fixed functions. 
To build this dataset, they first download every public GitHub event between March 2011 and October 2017 from GitHub Archive\footnote{https://www.gharchive.org/} and use the Google BigQuery APIs to identify all Java-file commits having a message containing the patterns \cite{fischer2003populating}: (“fix” or “solve”) and (“bug” or “issue” or “problem” or “error”). For each bug-fixing commit, they extract the source code before and after the fixing process by using the GitHub Compare API\footnote{https://developer.github.com/v3/repos/commits/\#compare-two-commits} to collect the buggy (pre-commit) and the fixed (post-commit) codes. Subsequently, they normalize all the names of the variables and custom methods, which greatly limits the vocabulary size and enables the model to focus on learning bug-fixing patterns. Then, they filter out the pairs that contain lexical or syntactic errors in either the buggy or fixed code, as well as the pairs with more than 100 atomic AST modification actions between the buggy and the fixed versions.
To achieve this, they employ the GumTree Spoon AST Diff tool \cite{falleri2014fine}. 
Finally, they divide the whole dataset into two subsets (\textit{small} with tokens $\leq 50$ and \textit{medium} with tokens $> 50$ and $\leq 100$) based on the code length. For the \textit{small} subset, the numbers of training, development, and test samples are 46,680, 5,835, and 5,835, respectively. For the \textit{medium} subset, the numbers are 52,364, 6,545, and 6,545, respectively.

\subsection{Text-to-code generation}
To carry out this task, we use CONCODE \cite{iyer2018mapping}, a widely used code generation dataset, which is collected from about 33,000 Java projects on GitHub. It contains 100,000 examples for training and 4,000 examples for validation and testing. Each example is a tuple consisting of NL descriptions, code environments and code snippets. The dataset is tasked with generating class member functions from natural language descriptions (Javadoc-style method comments) and class environments. Class environment is the programmatic context provided by the rest of the class, including other member variables and member functions in the class.

\subsection{Code summarization}
\label{Code_summarization_dataset}
For code summarization, we use the CodeSearchNet dataset \cite{husain2019codesearchnet}, which includes six programming languages; i.e., Python, Java, JavaScript, PHP, Ruby, and Go. The data comes from publicly available opensource non-fork GitHub repositories and each documentation is the first paragraph. 
We observe that some documents contain content unrelated to the function, such as a link ``http://..." that refers to external resources and an HTML image tag ``<img ...>" that inserts an image. 
Therefore, we filter the dataset  to improve its quality with the same four rules mentioned in Section \ref{code_search}.





The statistics about the filtered CodeSearchNet dataset used in CodeXGLUE are listed in Table \ref{table-codesearchnet-data-statistic}.

\begin{table}[h]
\begin{center}
\caption{Data statistics about the filtered CodeSearchNet dataset for the code summarization task.}
\label{table-codesearchnet-data-statistic}
		\begin{tabular}{lccc}
			\toprule
			Language & Training & Dev & Testing \\
			\midrule
			Go&167,288&7,325&8,122\\
			Java&164,923&5,183&10,955\\
			JavaScript&58,025&3,885&3,291\\
			PHP&241,241&12,982&14,014\\
			Python&251,820&13,914&14,918\\
			Ruby&24,927&1,400&1,261\\
			\bottomrule
		\end{tabular}
\end{center}
\end{table}

\begin{figure*}
    \centering
    \includegraphics[width=\textwidth]{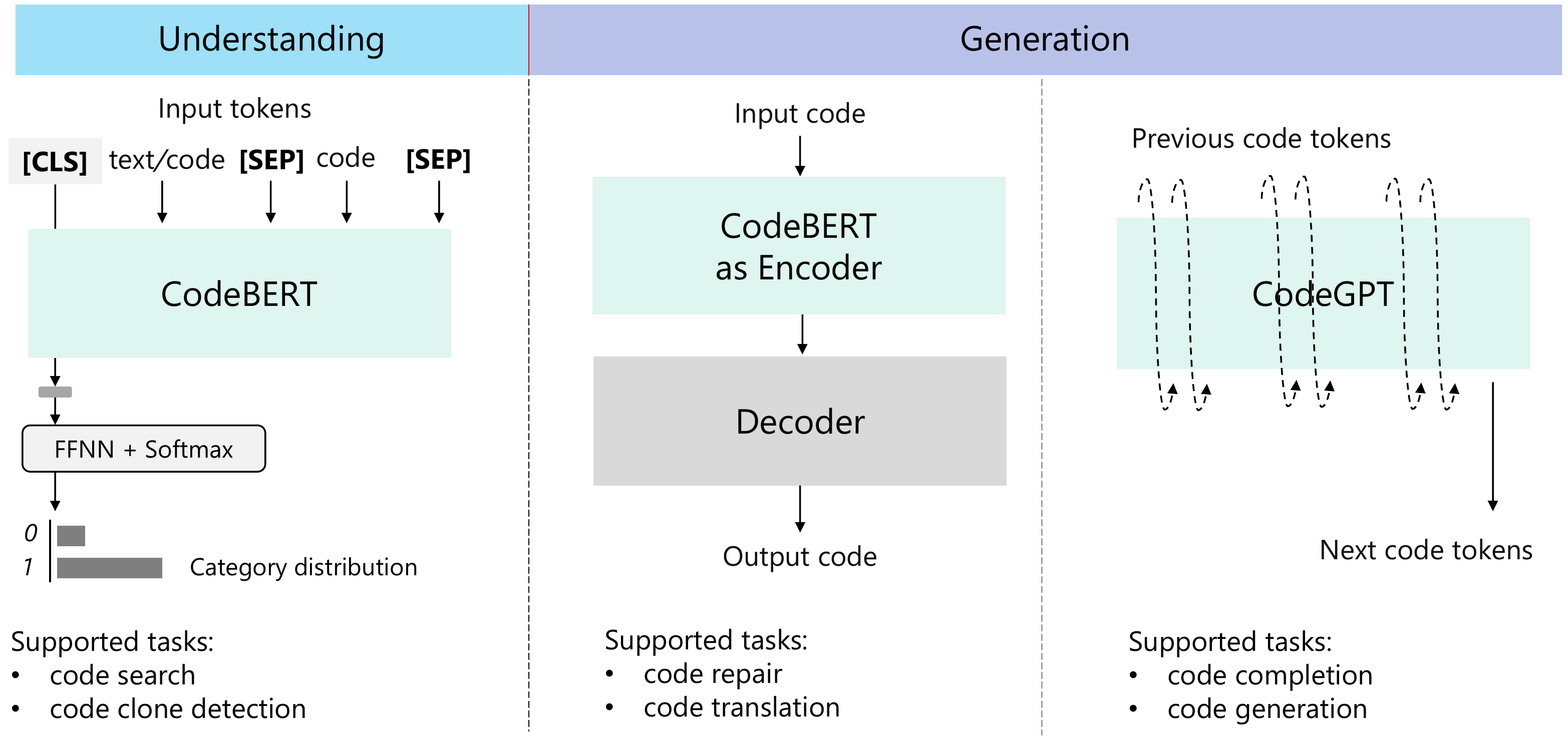}
    \caption{Three pipelines, including CodeBERT, CodeGPT, and Encoder-Decoder, are provided.}
    \label{fig:pipeline}
\end{figure*}

\subsection{Documentation translation}
\begin{figure}[t]
    \includegraphics[width=8.3cm]{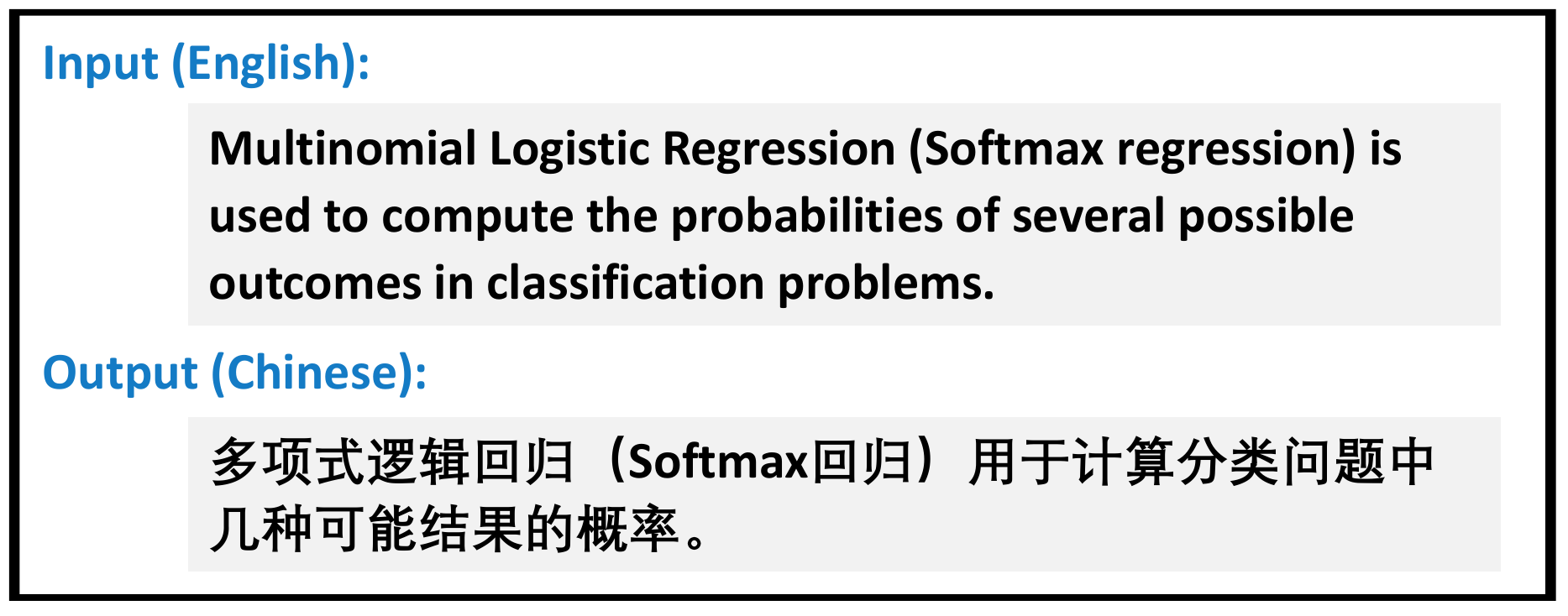}
    \caption{An English-to-Chinese example in the documentation translation dataset.}
    \label{fig:text2text}
\end{figure}

Documentation translation aims to translate code documentation automatically from one natural language (e.g., English) to another natural language (e.g., Chinese), as shown in Figure \ref{fig:text2text}.
The dataset we use is crawled  from Microsoft Documentation\footnote{https://docs.microsoft.com/, whose document is located at https://github.com/MicrosoftDocs/.}, including software and code description documents in different languages.
We focus on low-resource language pairs, where parallel
data is scarce, and introduce multilingual machine translation tasks, e.g., English $\Leftrightarrow$ Latvian, Danish, Norwegian, and Chinese.
To improve the data quality, we filter the corpus by removing the following examples.

(1) Pairs whose source sentence is the same as the target sentence;

(2) Pairs whose length of source language or target language is less than three words;

(3) Pairs whose length ratio between source and target languages is larger than three; 

(4) Pairs whose word alignment ratio computed by $\rm{fast\_align}$\footnote{https://github.com/clab/fast\_align.} is less than 0.6. 

The final
 training data includes 43K, 19K, 44K, and 50K sentence pairs for English $\Leftrightarrow$ Latvian, English $\Leftrightarrow$ Danish, English $\Leftrightarrow$ Norwegian, and English $\Leftrightarrow$ Chinese, respectively. In addition, each language pair has 1K development and test sentence pairs, respectively.

\section{Baseline Systems}
We provide three types of baseline models to perform the  previously mentioned tasks, including a BERT-style pretrained model (in this case, CodeBERT), which supports program understanding problems,   a GPT-style pretrained model called CodeGPT that helps us solve completion and generation problems, and an Encoder-Decoder framework that tackles sequence-to-sequence generation problems. 
An illustration of these three pipelines is shown in Figure \ref{fig:pipeline}. 

\subsection{CodeBERT}
To carry out code understanding tasks like clone detection, defect detection, cloze test, and code search, we use CodeBERT \cite{feng2020codebert} as our encoder. 
This is a bimodal pretrained model based on Transformer with 12 layers, 768 dimensional hidden states, and 12 attention heads for programming language (PL) and natural language (NL). 
\citet{feng2020codebert} pretrain CodeBERT by masked language modeling and replaced token detection objectives on the CodeSearchNet dataset \cite{husain2019codesearchnet}, which includes 2.4M functions with document pairs for six programming languages. 
The model supports different types of the sequence input like text/code and code/code with a special token  $[CLS]$ in front of the sequence and a special symbol $[SEP]$ to split two kinds of data types. 

The model is publicly available at \url{https://huggingface.co/microsoft/codebert-base}.

\subsection{CodeGPT}\label{section:codeGPT}
We provide CodeGPT, 
which is a Transformer-based language model pretrained on programming language (PL), to support the code completion and the text-to-code generation tasks. CodeGPT has the same model architecture and training objective of GPT-2 \cite{radford2019language}, which consists of 12 layers of Transformer decoders. More model settings are listed in Table \ref{table-model}. We pretrain monolingual models on Python and Java corpora from the CodeSearchNet dataset \cite{husain2019codesearchnet}, which includes 1.1M Python functions and 1.6M Java methods. Each function in training dataset has a function signature and a function body. Some functions also contain a natural language documentation. 

We train two CodeGPT models for each programming language. One model is pretrained from scratch, so that the BPE (byte pair encoder) \cite{sennrich2016bpe} vocabulary is newly obtained on the code corpus and the model parameters are randomly initialized. The other model is a domain-adaptive one, which uses the GPT-2 model as the starting point and is continually trained on the code corpus. As a result, the second model has the same GPT-2 vocabulary and natural language understanding ability. 
We refer to this model as \textbf{CodeGPT-adapted}, and regard it as the default one for the code completion and text-to-code generation tasks. 
Both models are publicly available at \url{https://huggingface.co/microsoft/CodeGPT-small-java} and \url{https://huggingface.co/microsoft/CodeGPT-small-java-adaptedGPT2}. \footnote{Replace "java" with "py" for models pre-trained on python dataset.}

\begin{table}[h]
\centering
\caption{Parameters of CodeBERT and CodeGPT models.}
\label{table-model}
\begin{tabular}{lcc}
    \toprule
    & CodeBERT & CodeGPT  \\
    \midrule
    Number of layers           & 12 & 12     \\
    Max length of position     & 512 & 1,024   \\
    Embedding size             & 768 & 768   \\
    Attention heads            & 12 & 12     \\
    Attention head size        & 64 & 64      \\
    Vocabulary size            & 50,265 & 50,000  \\
    Total number of parameters & 125M & 124M  \\
    \bottomrule 
\end{tabular}
\end{table}

\subsection{Encoder-Decoder}
For sequence-to-sequence generation problems like code repair, code translation, code summarization, and documentation translation, we provide an Encoder-Decoder framework. We initialize the encoder using CodeBERT \cite{feng2020codebert} and use a randomly initialized Transformer with 6 layers, 768 dimensional hidden states and 12 attention heads as the decoder in all settings.

\section{Experiment}
In this section, we report accuracy numbers of the baseline systems on 10 tasks. 
We will also show how long it takes to train the model and to do inference on the model.

\subsection{Clone Detection}
\paragraph{Setting} 
We use the \textbf{BigCloneBench} and \textbf{POJ-104} datasets for clone detection. 
The task of the \textbf{BigCloneBench} dataset is formulated as a binary classification to predict whether a given pair of codes has the same semantics, with the F1 score used as the evaluation metric. The task of the \textbf{POJ-104} dataset aims to retrieve 499 codes for a given code from the development/test set for validation/testing, with the Mean Average Precision (MAP) as the evaluation metric.  
The overall score of the clone detection task is the average value of F1 and MAP scores. 

\begin{table}[h]
\centering
\caption{Results on the clone detection task. 
}
\label{table-code-clone-detection}
\begin{tabular}{lccc}
\toprule
    & BigCloneBench & POJ-104& \\
    \midrule 
    Model& F1 & MAP& Overall\\
    \midrule
    RtvNN&1.0&-&- \\
    Deckard&3.0&-&- \\
    CDLH&82.0&-&- \\
    ASTNN &93.0&-&- \\
    FA-AST-GMN&95.0&-&- \\
    TBCCD & 95.0 & - & - \\
    \midrule
    code2vec*&-&1.98&- \\
    NCC*&-&54.19&- \\
    Aroma*&-&55.12&- \\
    MISIM-GNN* &-&82.45&- \\
	\midrule
	RoBERTa& 94.9&79.96&87.4 \\
	CodeBERT& \bf{96.5}&\bf{84.29}&\bf{90.4} \\
	\bottomrule
\end{tabular}
\end{table}

\begin{table*}
\caption{Results on the cloze test task.}
\label{table-clozetest}
\begin{center}
\begin{small}
\begin{tabular}{lcccccc|ccccccc}
\toprule
 \multirow{2}{*}{Model} & \multicolumn{6}{c}{CT-all} & \multicolumn{6}{c}{CT-maxmin} & \multirow{2}{*}{Overall} \\
 \cmidrule{2-13}
 & Ruby & JS & Go & Python & Java & PHP & Ruby & JS & Go & Python & Java & PHP & \\
\midrule
 RoBERTa & 47.44 & 59.96 & 40.77 & 54.35 & 50.73 & 60.16 
         & 73.68 & 64.71 & 71.71 & 59.18 & 59.75 & 69.78 
         & 62.45\\
 CodeBERT(MLM)  
                & {\bf 80.17} & {\bf 81.77} & \bf{83.31} & {\bf 87.21} & {\bf 80.63} & {\bf 85.05} 
                & {\bf 86.84} & {\bf 86.40} & \bf{90.79} & {\bf 82.20} & {\bf 90.46} & {\bf 88.21} 
                & {\bf 85.66}\\

\bottomrule
\end{tabular}
\end{small}
\end{center}
\end{table*}

\paragraph{Results}
Results achieved by different models are shown in Table \ref{table-code-clone-detection}. 
\textbf{RtvNN} \citep{white2016deep} trains a recursive autoencoder to learn representations for AST. 
\textbf{Deckard}  \citep{jiang2007deckard} computes vectors for structural information within ASTs and uses a Locality Sensitive Hashing (LSH) \citep{datar2004locality} to cluster similar vectors.
\textbf{CDLH} \citep{wei2017supervised} learns representations of code fragments via AST-based LSTM. 
\textbf{ASTNN} \cite{zhang2019novel} uses RNNs to encode AST subtrees for statements. It feeds the encodings of all statement trees into an RNN to learn representation for a program.
\textbf{FA-AST-GMN} \citep{wang2020detecting} uses GNNs over a flow-augmented AST to leverage explicit control and data flow information. 
\textbf{TBCCD} \citep{yu2019neural} proposes a position-aware character embedding and uses tree-based convolution to capture both the structural information of a code fragment from its AST and lexical information from code tokens.
\textbf{Code2vec} \cite{alon2019code2vec} learns representations of code snippets by aggregating multiple syntactic paths into a single vector. 
\textbf{NCC} \cite{ben2018neural} encodes programs by leveraging both the underlying data flow and control flow of the programs.
\textbf{Aroma} \cite{luan2019aroma} is a code recommendation engine that takes a partial code snippet and recommends a small set of succinct code snippets that contain the query snippet.
\textbf{MISIM-GNN} \cite{ye2020misim} learns a structural representation of code from context-aware semantic structure designed specifically to lift semantic meaning from the code syntax.

In this experiment, we use pretrained models like \textbf{RoBERTa} \cite{liu2019roberta} and \textbf{CodeBERT} \cite{feng2020codebert} to encode source code and take the representation to calculate semantic relevance of two codes through a feed forward
network or inner product.  Although CodeBERT does not leverage code structure that has proven to be effective in terms of code similarity measure  \cite{wei2017supervised,zhang2019novel,wang2020detecting,ben2018neural, ye2020misim}, the model still performs better than RoBERTa on the task of clone detection, achieving the overall score of 90.4. These experimental results demonstrate that pretraining is useful for clone detection.  
There is room for further improvement if code structure is further leveraged.

\subsection{Defect Detection}
\paragraph{Setting} We use the dataset mentioned in Section \ref{defect_detection_dataset} for defect detection, which aims to predict whether a source code contains defects that may be used to attack software systems. The  evaluation metric is accuracy score. We use the CodeBERT baseline to encode source code and take the representation of source code to calculate the probability of being exposed to vulnerabilities. 
\begin{table}[h]
\centering
\caption{Results on the defect detection task.}
\label{table-code-defect-detection}
\begin{tabular}{lc}
\toprule
    Model&Accuracy\\
    \midrule 
    BiLSTM&59.37\\
    TextCNN&60.69\\
    RoBERTa&61.05\\
    CodeBERT&\bf{62.08}\\
	\bottomrule 
\end{tabular}
\end{table}

\paragraph{Results} Table \ref{table-code-defect-detection} shows the results of the models we implemented. We use Bidirectional LTSM (\textbf{BiLTSM})  \cite{hochreiter1997long}, \textbf{TextCNN} \cite{kim2014convolutional}, \textbf{RoBERTa} \cite{liu2019roberta}, and \textbf{CodeBERT} \cite{feng2020codebert} to encode the representation of a source code, respectively. Then, a two-layer feed forward network followed by a softmax layer is used to calculate the probability of encountering vulnerabilities. 
As shown in the results, CodeBERT achieve a 62.1 accuracy score, resulting in state-of-the-art performance. However, the improvement achieved by the pretrained models is limited compared with \textbf{TextCNN}. A potential direction to improve these pretrained models is to incorporate information from code structures such as Abstract Syntax Tree, data flow, control flow, etc.


\subsection{Cloze test}
\paragraph{Setting} 
We use CT-all and CT-maxmin datasets for the cloze test task. Models are expected to predict the masked code token by leveraging documentation and the context of code.
Accuracy are reported for each language, with  the macro-average accuracy scores for all languages as the overall evaluation metric.

\paragraph{Results} Table \ref{table-clozetest} shows the results on the CT-all and CT-maxmin datasets. We report the performance of RoBERTa \cite{liu2019roberta} and CodeBERT (Masked Language Modeling, MLM) \cite{feng2020codebert}, which is initialized with RoBERTa and further trained with the masked language modeling objective. 
The results demonstrate that CodeBERT performs better than RoBERTa that only learns from natural language.

\subsection{Code completion}
\paragraph{Setting} 
We use the \textbf{PY150} and  \textbf{Github Java Corpus} datasets for token-level and line-level code completion tasks. The token-level task is to predict the next token given context of previous tokens, and predictions are evaluated according to token-level accuracy; whereas the line-level task entails the completion of a whole-line of code, and the quality of the code is evaluated through the metrics known as
exact match accuracy and Levenshtein edit similarity \cite{svyatkovskiy2020intellicode}. Levenshtein edit similarity measures how many single character edits are required to transform one string into another. This is a critical evaluation metric for the code completion scenario as it measures how much effort it takes for developers to correct an error in the code.
The score on each dataset is the average value of the accuracy on token-level completion and the edit similarity on line-level completion. The overall score of code completion task is calculated by averaging the scores on both datasets.

\begin{table*}[h]
    \centering
    \caption{Results on the code completion task.}
    \label{table-code-completion}
    \begin{tabular}{lccccccc}
    \toprule
    \multirow{3}{*}{Model} & \multicolumn{3}{c}{PY150} & \multicolumn{3}{c}{Github Java Corpus} & \multirow{3}{*}{Overall} \\
    \cmidrule{2-7}
    & token-level & \multicolumn{2}{c}{line-level} & token-level & \multicolumn{2}{c}{line-level} & \\
    \cmidrule{2-7}
    & Accuracy & EM & Edit Sim & Accuracy & EM & Edit Sim & \\
    \midrule
    LSTM & 58.00 & 17.93 & 50.05 & 56.02 & 10.30 & 41.55 & 51.41 \\
    Transformer & 73.26 & 36.65 & 67.51 & 64.16 & 15.33 & 50.39 & 63.83 \\
    GPT-2 & 74.22 & 38.55 & 68.94 & 74.89 & 24.30 & 60.70 & 69.69 \\
    CodeGPT & 74.93 & 39.11 & 69.69 & 76.45 & 25.30 & 61.54 & 70.65 \\
    CodeGPT-adapted & \bf{75.11} & \bf{39.65} & \bf{69.84} & \bf{77.13} & \bf{26.43} & \bf{63.03} & \bf{71.28} \\
    \bottomrule
    \end{tabular}
\end{table*}

\paragraph{Results} 
Table \ref{table-code-completion} shows the results of all models on both datasets. We fine-tune \textbf{LSTM} \cite{hochreiter1997long}, \textbf{Transformer} \cite{vaswani2017attention}, \textbf{GPT-2} \cite{radford2019language}, \textbf{CodeGPT} and \textbf{CodeGPT-adapted} to generate following tokens.
CodeGPT and CodeGPT-adapted models are described in Section \ref{section:codeGPT}.
CodeGPT-adapted achieves a state-of-the-art performance with the overall score of 71.28.

\subsection{Code search}
\paragraph{Setting} We use the \textbf{CodeSearchNet AdvTest} and \textbf{WebQueryTest} datasets mentioned in Section \ref{code_search} for code search. To improve efficiency, we separately encode text and code to perform code search. For the \textbf{CodeSearchNet AdvTest} dataset, the task is to find the most relevant code from a collection of candidates given a query and  it is evaluated through the Mean Reciprocal Rank (MRR) metric. For the \textbf{WebQueryTest} dataset, the task is formulated as a binary classification to predict whether a code can answer a given query
and we use the F1 and accuracy scores as evaluation metrics. The overall score for code search is the average of the values recorded for the two subtasks. 

\paragraph{Results} Table \ref{table-code-search-results} presents the results on the CodeSearchNet AdvTest and WebQueryTest datasets. We report the performance of RoBERTa \cite{liu2019roberta} and CodeBERT \cite{feng2020codebert}. The table shows that CodeBERT performs better than RoBERTa.

\begin{table}[h]
\centering
\caption{Results on the code search task.}
\label{table-code-search-results}
\begin{tabular}{lcccc}
\toprule
    & AdvTest & \multicolumn{2}{c}{WebQueryTest}& \\
    \midrule
     Model & MRR &  F1 & Accuracy & Overall \\
    \midrule 
	RoBERTa& 18.33 &  57.49 & 40.92 & 33.63 \\
	CodeBERT& \bf{27.19} & \bf{58.95} & \bf{47.80} & \bf{40.28} \\
	\bottomrule
\end{tabular}
\end{table}

\subsection{Text-to-code generation}
\paragraph{Setting}
We use the CONCODE dataset for text-to-code generation. Models are expected to generate source codes of Java class member functions, given natural language descriptions and class environments. We report the exact match accuracy, the BLEU score \cite{papineni2002bleu}, and the CodeBLEU score \cite{ren2020codebleu}. We use the CodeBLEU score as the overall evaluation metric.

\paragraph{Results}
Table \ref{table-text-to-code} presents the results on the CONCODE test set. \textbf{Seq2Seq} \cite{sutskever2014sequence} is an RNN-based sequence to sequence model. \textbf{Seq2Action + MAML} \cite{guo2019coupling} combines a context-aware retrieval model with model-agnostic meta-learning (MAML). \textbf{Iyer-Simp + 200 idoms} \cite{iyer2019learning} extracts code idioms and applies idioms-based decoding. We also report the performance of pretrained models, including \textbf{GPT-2} \cite{radford2019language}, \textbf{CodeGPT}, and \textbf{CodeGPT-adapted}. CodeGPT-adapted achieves the CodeBLEU score of 35.98, resulting in a state-of-the-art performance.

\begin{table}[h]
    \centering
    \caption{Results on the text-to-code generation task.}
    \label{table-text-to-code}
    \begin{tabular}{lccc}
    \toprule
    Model & EM & BLEU & CodeBLEU\\
    \midrule
    Seq2Seq & 3.05 & 21.31 & 26.39 \\
    Seq2Action+MAML & 10.05 & 24.40 & 29.46 \\
    Iyer-Simp+200 idoms & 12.20 & 26.60 & - \\
    GPT-2 & 17.35 & 25.37 & 29.69 \\
    CodeGPT & 18.25 & 28.69 & 32.71 \\
    CodeGPT-adapted & \bf{20.10} & \bf{32.79} & \bf{35.98} \\
    \bottomrule
    \end{tabular}
\end{table}

\begin{table*}
    \centering
    \caption{Results on the code repair task.}
    \label{code-repair-result}
    \begin{tabular}{lccccccc}
    \toprule
        \multirow{2}*{Method} & \multicolumn{3}{c}{small} & \multicolumn{3}{c}{medium} & \multirow{2}*{Overall}\\
        \cmidrule{2-7}
        & BLEU & Acc & CodeBLEU & BLEU & Acc & CodeBLEU\\
        \midrule
        Naive & \textbf{78.06} & 0.000 & - & 90.91 & 0.000 & - & 0.000\\
        LSTM & 76.76 & 0.100 & - &  72.08 & 0.025 & - & 0.063\\
        Transformer & 77.21 & 0.147 & 73.31 & 89.25 & 0.037 & 81.72 & 0.092 \\
		CodeBERT & 77.42 & \textbf{0.164} & \textbf{75.58} & \textbf{91.07} & \textbf{0.052} & \textbf{87.52} & \textbf{0.108} \\
		\bottomrule
        \end{tabular}
\end{table*}

\subsection{Code translation}
\paragraph{Setting}
We use the dataset we build as described in Section \ref{code_translation}. The dataset contains matching samples of Java and C\# functions. We report the exact match accuracy, the BLEU score \cite{papineni2002bleu} and the CodeBLEU score \cite{ren2020codebleu} on this task. CodeBLEU is used as the overall evaluation metric. 

\paragraph{Results}
Table \ref{code-translation-result} shows the results of models on both translation directions. The \textbf{Naive} method directly copies the source code as the translation result. \textbf{PBSMT} is short for phrase-based statistical machine translation \citep{koehn2003statistical}. \textbf{Transformer} uses the same number of layers and hidden size as the pretrained models. The table shows that Transformer initialized with CodeBERT and fine-tuned with the matching sample pairs produces the best result.

\begin{table*}
    \centering
    \caption{Results on the code translation task.}
    \label{code-translation-result}
    \begin{tabular}{lccccccc}
        \toprule
        \multirow{2}*{Method} & \multicolumn{3}{c}{Java$\to$C\#} & \multicolumn{3}{c}{C\#$\to$Java} & \multirow{2}*{Overall}\\
        \cmidrule{2-7}
        & BLEU & Acc & CodeBLEU & BLEU & Acc & CodeBLEU\\
        \midrule
        Naive & 18.54 & 0.000 & - & 18.69 & 0.000 & - & -\\
        PBSMT & 43.53 & 0.125 & 42.71 & 40.06 & 0.161 & 43.48 & 43.10 \\
        Transformer & 55.84 & 0.330 & 63.74 & 50.47 & 0.379 & 61.59 & 62.67 \\
    	RoBERTa (code) & 77.46 & 0.561 & 83.07 & 71.99 & 0.579 & \textbf{80.18} & 81.63 \\
		CodeBERT & \textbf{79.92} & \textbf{0.590} & \textbf{85.10} & \textbf{72.14} & \textbf{0.580} & 79.41 & \textbf{82.26} \\
		\bottomrule
        \end{tabular}
\end{table*}

\subsection{Code repair}
\paragraph{Setting}
We use the dataset originally released by \citet{tufano2019empirical}, which is described in Section \ref{code_repair}. The dataset contains two subsets established according to the length of the Java functions: \textit{small} $\leq 50$ and $50 <$ \textit{medium} $\leq 100$ . We report the exact match accuracy, the BLEU score \cite{papineni2002bleu} and the CodeBLEU score \cite{ren2020codebleu} on this task. The exact match accuracy is used as the overall evaluation metric. 

\paragraph{Results}
The \textbf{Naive} method  directly copies the buggy code as the repair result. As for \textbf{Transformer}, we use the same number of layers and hidden size as the pretrained models. 
With regard to the \textbf{CodeBERT} method, 
we initialize the Transformer encoder with pretrained CodeBERT model and randomly initialize the parameters of the decoder and the source-to-target attention. Then we use the training data to fine-tune the whole model. As shown in the table, Transformer with CodeBERT initialization achieves the best performance among all models.

\subsection{Code Summarization}
\paragraph{Setting}
We use the dataset mentioned in Section \ref{Code_summarization_dataset} for code summarization. To evaluate the models, we follow \citet{feng2020codebert}, who use smoothed BLEU score \cite{lin2004orange} as evaluation metric, because this is suitable for evaluating short documents. 
We use the encoder-decoder pipeline to tackle this problem.
The max length of input and inference are set as 256 and 128, respectively. We use
the Adam optimizer to update the models' parameters. The learning rate and the batch size are 5e-5 and 32, respectively. We tune the hyperparameters and perform early stopping on the development set. 
\begin{table*}
\begin{center}
\caption{Results on the code summarization task.}
\label{table-code2nl-reults-codesearchnet}
\begin{tabular}{lccccccc}
\toprule
Model & Ruby & Javascript & Go & Python & Java & PHP & Overall\\
\midrule
Seq2Seq & 9.64 & 10.21 & 13.98 & 15.93 & 15.09 & 21.08 & 14.32\\
Transformer & 11.18 & 11.59 & 16.38 & 15.81 & 16.26 & 22.12 & 15.56\\
RoBERTa & 11.17 & 11.90 & 17.72 & 18.14 & 16.47 & 24.02 & 16.57\\
CodeBERT  & {\bf 12.16} & {\bf 14.90} & \bf{18.07} & {\bf 19.06} & {\bf 17.65} & {\bf 25.16} & {\bf 17.83}\\
\bottomrule
\end{tabular}

\end{center}
\end{table*}

\paragraph{Results} Table \ref{table-code2nl-reults-codesearchnet} shows the results achieved by different models in code summarization. \textbf{Seq2Seq} is an RNN-based sequence to sequence model. \textbf{Transformer} and \textbf{RoBERTa} use the same setting as \textbf{CodeBERT}, but the encoder is initialized randomly and by RoBERTa \cite{liu2019roberta}, respectively. All models use Byte Pair Encoding (BPE) \cite{sennrich2015neural} vocabulary. In this experiment, CodeBERT obtains a 1.3\% gain in the BLEU score over RoBERTa and achieves the state-of-the-art performance on six programming languages.

\subsection{Documentation translation}
\paragraph{Setting}
We use the Microsoft Docs dataset for text-to-text translation tasks, which focus on low-resource multilingual translation between English (EN) and other languages, including Latvian (LA), Danish (DA), Norwegian (NO), and Chinese (ZH).
Following \citet{johnson2017google}, we train a single multilingual model as our baseline.
To distinguish between different translation pairs, we add an language token (e.g., $\langle\mathit{2en}\rangle$, $\langle\mathit{2zh}\rangle$) at the beginning of the source sentence to indicate the target language the model should translate.
We  initialize the encoder of the multilingual translation model with XLM-R \cite{conneau2019unsupervised}.
Models are evaluated through BLEU \cite{papineni2002bleu} score,
and the overall score for documentation translation is the average BLEU score on the eight translation directions.

\begin{figure*}
    \centering
    \includegraphics[width=0.97\textwidth]{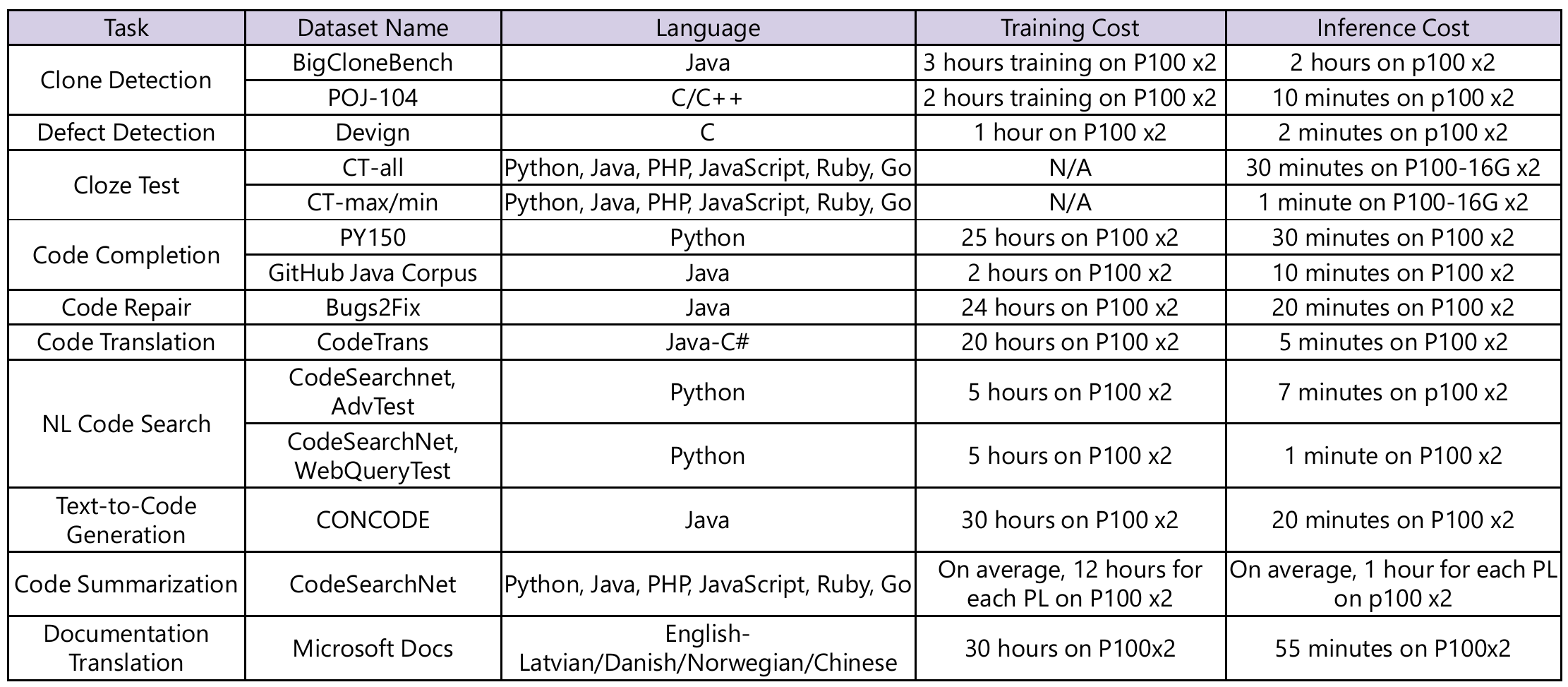}
    \caption{Training and inference time costs for each task, evaluated on two P100 GPUs.}
    \Description{Training and inference time costs.}
    \label{fig:time}
\end{figure*}

\paragraph{Results}
Table \ref{documentation-translation-result} shows the results achieved by the models on eight translation directions.
\textbf{Transformer Baseline} is the multilingual translation model \cite{johnson2017google}.
\textbf{pretrained Transformer} initializes the encoder of \textbf{ Transformer Baseline} with XLM-R\cite{conneau2019unsupervised}.
In terms of overall performance on the eight translation directions, \textbf{Transformer Baseline} and \textbf{pretrained Transformer} obtain the BLEU score of 52.67 and 66.16, respectively.
Experimental results demonstrate that pretraining achieves a 13.49 improvement in BLEU score over strong baseline model. Figure \ref{fig:time} shows how long it takes to train the model and to do inference on the model, as well as in other tasks.

\begin{table}[t]
	\begin{center}
	\caption{Results on the documentation translation task.}
		\begin{tabular}{lp{2cm}<{\centering}p{2cm}<{\centering}} 
			 \toprule 
			Task & Transformer Baseline & pretrained
			Transformer \\
			 \midrule
			EN $\rightarrow$ DA  & 53.31     &  \textbf{67.09}  \\
			EN $\rightarrow$ LA  & 37.85     &  \textbf{51.92}  \\
			EN $\rightarrow$ NO  & 53.84    &  \textbf{68.00}  \\
			EN $\rightarrow$ ZH  & 59.90     &  \textbf{70.60}  \\
			 \midrule
			DA $\rightarrow$ EN  & 58.73     &  \textbf{67.02}  \\
			LA $\rightarrow$ EN  & 50.37     &  \textbf{68.30}  \\
			NO $\rightarrow$ EN  & 57.73     &  \textbf{71.84}  \\
			ZH $\rightarrow$ EN  & 50.00     &  \textbf{64.47}  \\
			 \midrule
			Overall    & 52.67     &  \textbf{66.16}  \\
			\bottomrule
		\end{tabular}
	\label{documentation-translation-result}
	\end{center}
\end{table}

\section{Related Work}
Benchmark datasets have been playing a central role in the growth of applied AI research. 
For example, the LibriSpeech \cite{panayotov2015librispeech} 
and the SQuAD \cite{rajpurkar2016squad} datasets
drive the development of data-driven models for automatic speech recognition and reading comprehension of text, respectively.
With the growing demand for testing models' generalization ability on a wide range of applications, researchers have created or assembled datasets that cover many tasks. Representative samples of these datasets include ImageNet \cite{deng2009imagenet} for computer vision, GLUE \cite{wang2018glue} for natural language understanding, XTREME \cite{hu2020xtreme} and XGLUE \cite{liang2020xglue} for cross-lingual natural language processing. To the best of our knowledge, CodeXGLUE is the first diversified benchmark dataset that can be applied to various code intelligence problems.

Many tasks related to machine learning for software engineering \cite{allamanis2018survey} have sufficient amount of data to support the development of data-driven methods, but are not covered by CodeXGLUE.
We plan to extend to these tasks in the future. 
%
For example,
the {idiom mining task} \cite{allamanis2014mining,iyer2019learning} is to extract code idioms, which are syntactic fragments that recur across software projects and serve a single semantic purpose \cite{allamanis2014mining}. 
{Bug localization} \cite{ray2016naturalness,gupta2019neural,vasic2019neural} is to point the error location when a program fails tests. 
The {test case generation} task \cite{fraser2011evosuite,tufano2020unit} is to generate unit test cases automatically. 
The {program synthesis}  \cite{neelakantan2015neural,reed2015neural,singh2015predicting,vijayaraghavan2017bayesian,feser2015syn,kulal2019spoc,zhong2020semantic} extends the text-to-code generation task aims to generate programs from a specification \citep{gulwani2017program}, such as pseudocode, natural language description,  and input/output examples. 

\section{Conclusion}
With CodeXGLUE, we seek to support the development of models that can be applied to various program understanding and generation problems, with the goal of increasing the productivity of software developers. We encourage researchers to participate in the open challenge to make progress in code intelligence. Moving forward, we are planning to extend CodeXGLUE to more programming languages and downstream tasks while continuing to develop advanced pretrained models by exploring new model structures, introducing new pretraining tasks, using different types of data, and more.



\bibliographystyle{ACM-Reference-Format}
\bibliography{arxiv}

\appendix

\end{document}